\documentclass[12pt]{article}
\newcommand\be{\begin{equation}}
\newcommand\ee{\end{equation}}
\newcommand\ba{\begin{eqnarray}}
\newcommand\ea{\end{eqnarray}}
\newcommand\eq{\begin{equation}}           
\newcommand\en{\end{equation}}

\input epsf

\usepackage{amssymb}

\usepackage{epsfig,cite}
\include{epsf}

\usepackage{subfigure}

\begin{document}
\title{
{\hfill  \small MCTP-09-07,FTPI-MINN-09/07, UMN-TH-2737/09  \\ ~\\~\\}
A Sneutrino NLSP in the $\nu$CMSSM
}
\author{Kenji Kadota$^{1,2}$, Keith A. Olive$^2$ and Liliana Velasco-Sevilla$^3$ \\
$^1$ {\em \small Michigan Center for Theoretical Physics, University of Michigan, Ann Arbor, MI 48109} \\
$^2$ {\em \small William I. Fine Theoretical Physics Institute, University of Minnesota, Minneapolis, MN 55455} \\
$^3$ {\em \small The Abdus Salam International Center for Theoretical Physics, 34014 Trieste, Italy} 
}
\maketitle   

\begin{abstract}
We extend the constrained minimal supersymmetric model (CMSSM) by adding a right-handed neutrino superfield ($\nu$CMSSM) which decouples close to the GUT scale. We study the effects of a right-handed neutrino on the low energy spectrum and focus on the predictions for dark matter properties. 
We pay particular attention to the realization of the light (left-handed) sneutrino which can be 
the next-to-lightest supersymmetric particle (NLSP) with either the neutralino or gravitino as the
lightest supersymmetric particle (LSP). Notably, for the case of a neutralino LSP with a sneutrino NLSP, there are new  `sneutrino coannihilation regions' which yield the desired thermal neutralino relic density determined by WMAP.  
\end{abstract}

\setcounter{footnote}{0} 
\setcounter{page}{1}
\setcounter{section}{0} \setcounter{subsection}{0}
\setcounter{subsubsection}{0}

\section{Introduction}

One explanation for the tiny masses of left-handed neutrinos in the Standard Model is the seesaw mechanism \cite{seesaw}. This is accomplished through
the introduction of a heavy right-handed neutrino. 
In this letter, we extend the constrained minimal supersymmetric model (CMSSM) by adding a right-handed neutrino superfield and study how such a gauge singlet field with a Majorana mass close to the grand unification (GUT) scale can affect low-energy phenomenology.  The effects of a GUT scale right-handed neutrino in the supersymmetric version of the seesaw mechanism  on lepton flavor violation and gauge unification have been studied 
\cite{hisa,casa,how4,baer2,bla,hitoshi2,gordy3,hints,de,iba,hir}. Differences between this model and the 
more commonly studied CMSSM \cite{funnel,cmssm,efgosi,eoss,cmssmwmap}  arise due to 
changes in the renormalization group equation (RGE) evolution.
Because the mass spectra at the low energy can be affected, the shape and position of the supersymmetric (SUSY) parameter regions which lead 
to the desirable dark matter relic abundance can change accordingly \cite{petcov,cali,barg,gom}. 
In many cases, the change in the low energy spectrum simply results in a shift of the CMSSM
parameters which is necessary to obtain the correct relic density. Here, we study the effects of a heavy right-handed neutrino on dark matter properties and pay particular attention to the role of the light left-handed sneutrinos. We focus on a new `sneutrino coannihilation region' which is found in the 
$\nu$CMSSM. This is a distinctive region of parameter space where the characteristic physical processes involved which lead to the desired LSP neutralino thermal relic abundance (i.e. sneutrino annihilation and coannihilation) are different from those in the conventionally studied regions in CMSSM, namely, regions defined by stau coannihilations, stop coannihilations, the A-pole funnel, focus-point or hyperbolic branch and bulk regions \cite{stau,stop,funnel,efgosi,focus}.

As we will see, there are regions of the $\nu$CMSSM parameter space in which the sneutrino
mass is driven close to and sometimes below the lightest neutralino mass. 
In principle, the sneutrino is a possible candidate for dark matter. However, its relic density is
sufficiently high only for relatively large sneutrino masses ($\gtrsim 500$ GeV) and even then
direct detection experiments place strong constraints on the abundance of sneutrinos \cite{directleft}
\footnote{We note that this conclusion can be avoided in some models with lepton number violation \cite{hmm}.}. As a consequence, we will consider models where the sneutrino is the NSLP,
with $m_\chi < m_{\tilde \nu}$.  If this condition is violated, we will in addition consider
models with a gravitino LSP.

Light left-handed sneutrinos are also possible in models with non-universal soft SUSY breaking scalar mass, such as the non-universal Higgs mass model (NUHM) \cite{nonu,nuhm,nuhm1,eosk4}, or in models where the gravitino is the LSP with a sneutrino NLSP \cite{roeck,buch,covi2,laura,james,john2}. A sneutrino NLSP in the $\nu$CMSSM can be realized with either a  gravitino LSP or a neutralino LSP, using only the universal soft-SUSY breaking parameters (including those of the right-handed neutrino) and does not rely on any departures from universality.  

We first present the model and a short description of our numerical analysis procedures in \S \ref{mode}. Our main results,  the realization of the sneutrino coannihilation regions in the $\nu$CMSSM are given in \S \ref{snlsp} followed by concluding remarks.

\section{The $\nu$CMSSM Model}
\label{mode}
For simplicity we illustrate our results in a concrete example which includes one heavy right-handed neutrino superfield $N$ which we associate with the third generation, i.e., we are ignoring flavor mixings in the neutrino sector. In this case, it is the tau sneutrino which can become much lighter relative to its mass in the CMSSM. Our superpotential is
\begin{equation}
W=W_{MSSM}+y_{N} {L} {H}_u {N}^c
+{1\over 2}M_N {N}^c {N}^c 
\label{sprhn}
\end{equation}
where $W_{MSSM}$ is the standard MSSM superpotential. We assume that the soft SUSY breaking parameters for the right-handed neutrino, such as  the soft SUSY breaking scalar mass $m_{N}$ and the trilinear coupling $A_N$, also share the universal values in CMSSM at the GUT scale $Q_{GUT}$. Hence, in addition to the conventional CMSSM parameters represented by  the universal scalar mass, gaugino mass, trilinear mass, ratio of the Higgs vevs, and the sign of the $\mu$ parameter, 
\ba
m_0,M_{1/2},A_0,\tan \beta, \mbox{sign}(\mu), 
\ea
the $\nu$CMSSM has two more input parameters which we take to be
\ba
\label{nmass}
M_N(Q_{GUT}), m_{\nu}(Q_{M_Z})
\ea
We specify the right-handed neutrino mass $M_N$ at the GUT scale $Q_{GUT}$ \footnote{We are interested in the parameter range $M_N\leq Q_{GUT}$.}, and fix the left-handed neutrino mass at the scale of Z boson mass $Q_{M_Z}$. Note that the dominant contributions to the mass of the right handed sneutrino $M_{\tilde{N}}$ comes from the heavy neutrino mass $M_N$ rather than from the electroweak scale soft SUSY breaking mass $m_{{N}}$, so that $N$ and $\tilde{N}$ are decoupled at the energy scale $M_N$.

We numerically evolve the full two-loop RGEs \cite{baer2,iba,ant,ant2} including a right-handed neutrino from the GUT scale $Q_{GUT}$ (which is obtained iteratively assuming gauge coupling unification, typically $Q_{GUT}\sim 2 \times 10^{16}GeV$) down to 
the right-handed neutrino mass scale. Below $M_N$, the heavy right-handed neutrino is integrated out and the 
consequent non-renormalizable terms in the Lagrangian are suppressed by $M_N$. These 
non-renormalizable terms do not have the significant effects on the mass spectrum at the electroweak scale, with the possible exception of the light left-handed neutrino mass which receives its only contribution from the dimension five operator 
\ba
\label{5di}
L_5 \ni -\kappa (L H_u) (LH_u)
\ea
Hence, the light neutrino mass at the electroweak scale is given by $m_{\nu}(Q_{M_Z})=\kappa \langle H_u \rangle ^2$ and is taken to be one of our input parameters. We included the running of the dimensionful parameter $\kappa$ in our numerical treatment of RGEs below $M_N$ by matching those RGE solutions accordingly at $M_N$ \cite{ant,ant2}.

\section{A Sneutrino NLSP in the $\nu$CMSSM}
\label{snlsp}
\begin{figure}[ht]
\begin{center}    
\epsfxsize = 0.5\textwidth
\epsffile{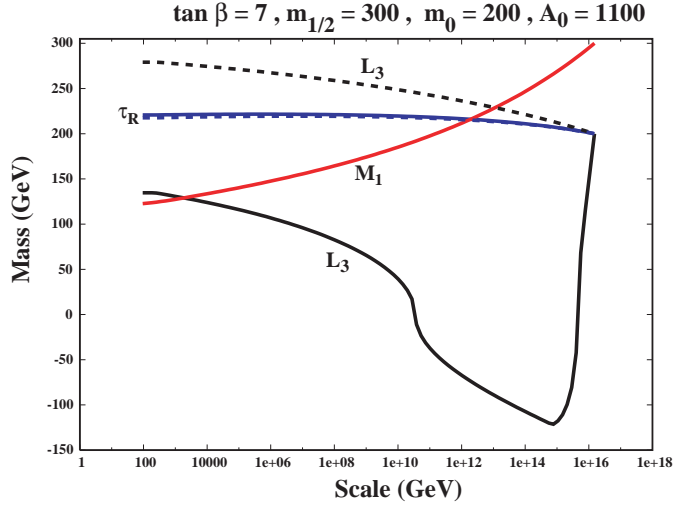}
\end{center}        
\caption{
\label{rge} \it{
The RGE evolution of the soft mass parameters corresponding to $L_3$, $\tau_R$, and the gaugino mass, $M_1$ as a function of renormalization scale. Initial conditions are set at $m_{L_3} = m_{\tau_R} = 200$ GeV and $M_1 =  300$ GeV. The CMSSM evolution is shown by dashed curves while the $\nu$CMSSM is shown by solid curves.  In the latter, $M_N = 10^{15}$ GeV, with $m_\nu = 0.05$ eV. 
The gaugino mass and right-handed slepton are affected at the two-loop
level by a right-handed neutrino and these effects are not visible in this
figure. Negative values refer to the sign 
of the mass-squared values.}}
\end{figure}

In the CMSSM, right-handed charged sleptons are in general lighter than the corresponding left-handed states because the beta functions of the latter receive contributions from $SU(2)_L$ weak couplings in addition to U(1) hypercharge couplings. This evolution is seen by the dashed curves in 
Fig. \ref{rge} where we show the 
running of the soft mass parameters associated with the left-handed lepton doublet, $L_3$,
the right-handed state, $\tau_R$, along with the U(1) gaugino mass, $M_1$. Starting with a common
scalar mass at the GUT scale, we see the left-handed mass parameter running to higher values.
Note the mass eigenstates  of the sleptons will be further split by L-R mixing terms
proportional to the charged lepton mass. For this reason, the lighter tau slepton is mostly composed of a 
right-handed component and is often the NLSP in CMSSM models.

This conventional picture changes with the introduction of a heavy right-handed Majorana neutrino.
We can gain insight to the qualitative behavior of the running by following the 1-loop contributions in the RGEs for the squared masses of the slepton doublet and the charged right-handed slepton 
\ba
\frac{d}{dt}m_{L_3}^2&=&\frac{1}{16 \pi^2}\left(2 y_{\tau}^2X_{\tau}+2 y_N^2X_N-\frac65g_1^2 M_1^2-6g_2^2 M_2^2-\frac{3}{5}g_1^2 S\right)+... \label{rge1} \\
\frac{d}{dt}m_{\tau_R}^2&=& \frac{1}{16 \pi^2}\left(4 y_{\tau}^2X_{\tau}-\frac{24}{5}g_1^2 M_1^2+\frac{6}{5}g_1^2 S \right)+...
\ea
where the $y$'s are Yukawa couplings, $g$'s are gauge couplings,  $t=\log Q$ and 
\ba
X_\tau&=&m_{L_3}^2 + m_{\tau_R}^2 + m_{H_d}^2 + A_\tau^2\\
X_N&=&m_{L_3}^2+m_{{N}}^2+m_{H_u}^2+A_N^2 \\
S&=&Tr(m_{ Q}^2 + m_{ D}^2 -2 m^2_{ U}
-m_{ L}^2 +m_{E}^2)
+ {m}^2_{H_u}-{m}^2_{H_d}
\ea
In the CMSSM (as well as the $\nu$CMSSM) $S=0$ at the GUT scale due to the universality of the soft scalar masses. Deviations from $S=0$ are due to the RGE evolution at the two-loop level. Hence, $S$ does not play a significant role for our study. This is in contrast to the NUHM \cite{nuhm},
where the non-universality of the Higgs masses allows for non-zero initial values for $S$. 
For large negative contributions to $S$ (eg. when $m_{H_d}^2 > m_{H_u}^2$) at the GUT scale, 
the left-handed states will be driven to values lower than the right-handed state allowing for the 
possibility for relatively light sneutrinos \cite{john2}. In the $\nu$CMSSM, there is a new
term in the above RGEs: $2 y_N^2 X_N$, which will effectively play the role of a negative contribution to $S$.  When the heavy right handed neutrino mass scale $M_N$ is close to the GUT scale, $y_N$ 
will be large and $2 y_N^2 X_N$ can affect the REG evolution of $m^2_{L_3}$ significantly to make $m^2_{L3}$ smaller than $m^2_{\tau_R}$. This is shown by the solid curves in Fig. \ref{rge}. This is similar to point C studied in Ref. \cite{barg}.
Thus, for certain regions in the $\nu$CMSSM parameter space, the sneutrino may be close to or even be below the neutralino mass.  If $m_{\tilde \nu} > m_\chi$, we could have a neutralino LSP with a sneutrino NLSP.  When $m_{\tilde \nu} < m_\chi$, we will assume a gravitino LSP and sneutrino NLSP for reasons discussed above. We will focus on the case of the neutralino LSP as the `sneutrino coannihilation region' in this context has not been explored before. We will later follow with a brief discussion on the gravitino LSP scenario.

We next discuss the parameter choices necessary for the realization of a sneutrino NLSP\footnote{We restrict the following discussion to $\mu > 0$.}. 
First, it is clear that a  small universal scalar mass $m_0$ is preferred to keep the sneutrino mass light
as every scalar mass is affected by $m_0$. To drive $m_{L_3}^2$ to low values, we need large $X_N$ and so we should choose a relatively large value of $A_0$. As we will see, this will also lead to an increased
mass for the light Higgs scalar. 
 A small value for $m_{1/2}$ is also preferred because the left-handed soft masses are more sensitive to $m_{1/2}$ than are their right-handed counterparts. In other words, we will expect that
 the mostly right-handed stau eigenstate will 
become lighter than the sneutrino at large $m_{1/2}$. 
Finally, a moderate value for $\tan \beta$ should be chosen so that left-right mixing in the stau mass matrix does not push one of the stau eigenstates below the sneutrino mass. 
We note that for those choices of relatively small $\tan \beta, m_{1/2}$ and $m_0$ there is in general a tight constraint from the lower bound on the Higgs mass, $m_h>114.4$ GeV \cite{lepa}. However, our choice of  relatively large $A_0$ will help alleviate this problem in part because a large positive value for $A_0$ can increase the left-right mixing in the stop mass matrix which can enhance the stop loop contributions to the Higgs mass. Choosing $A_0$ excessively large, however,  causes the Higgs mass to decrease and also leads to tachyonic squark and slepton masses.

\begin{figure}
             \centering
             \mbox{
              \subfigure[CMSSM] 
             {
                 \label{am1}
                 \includegraphics[width=0.47\textwidth]{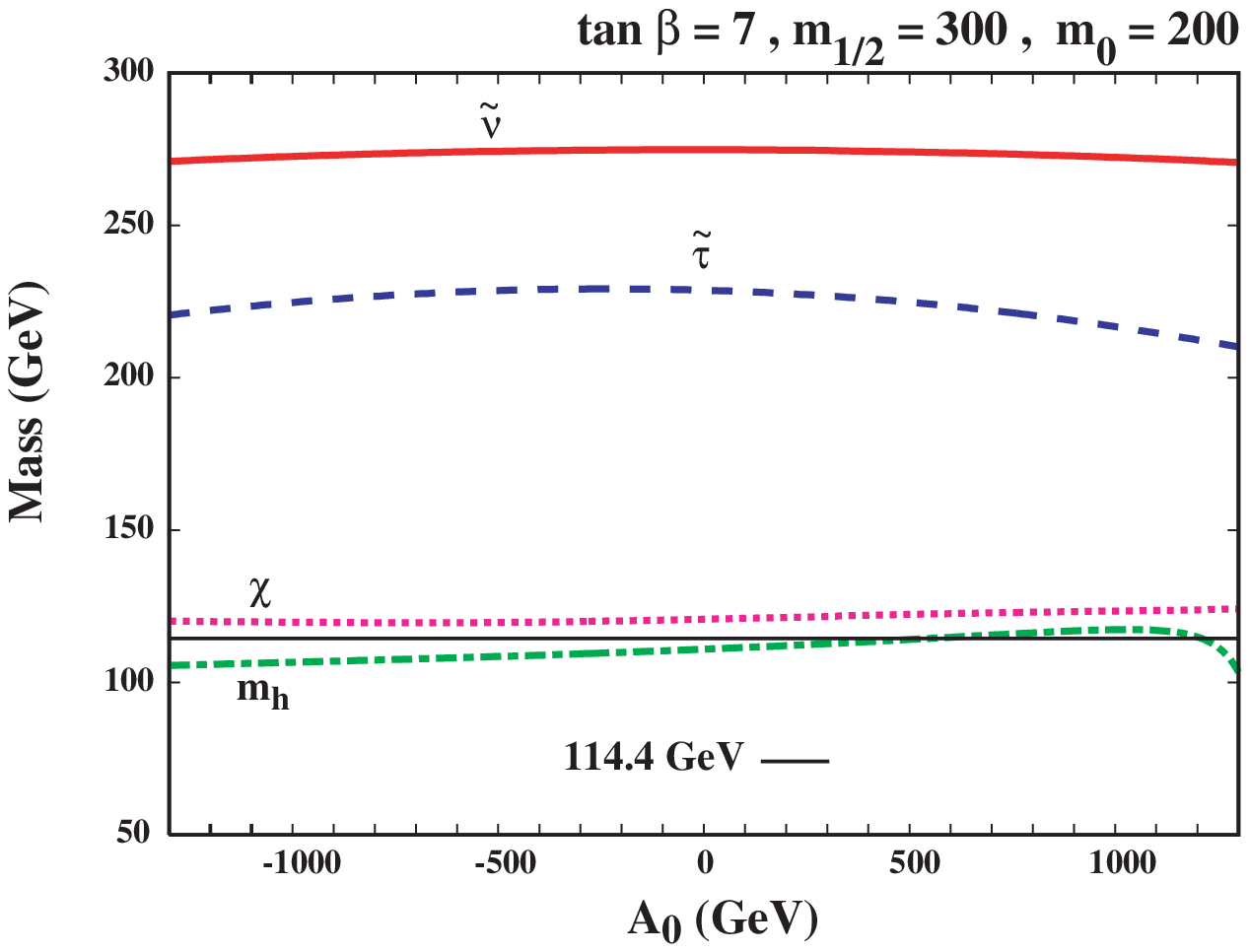}
             }
             \quad
            \subfigure[$\nu$CMSSM] 
             {
               \label{am11}
                 \includegraphics[width=0.47\textwidth]{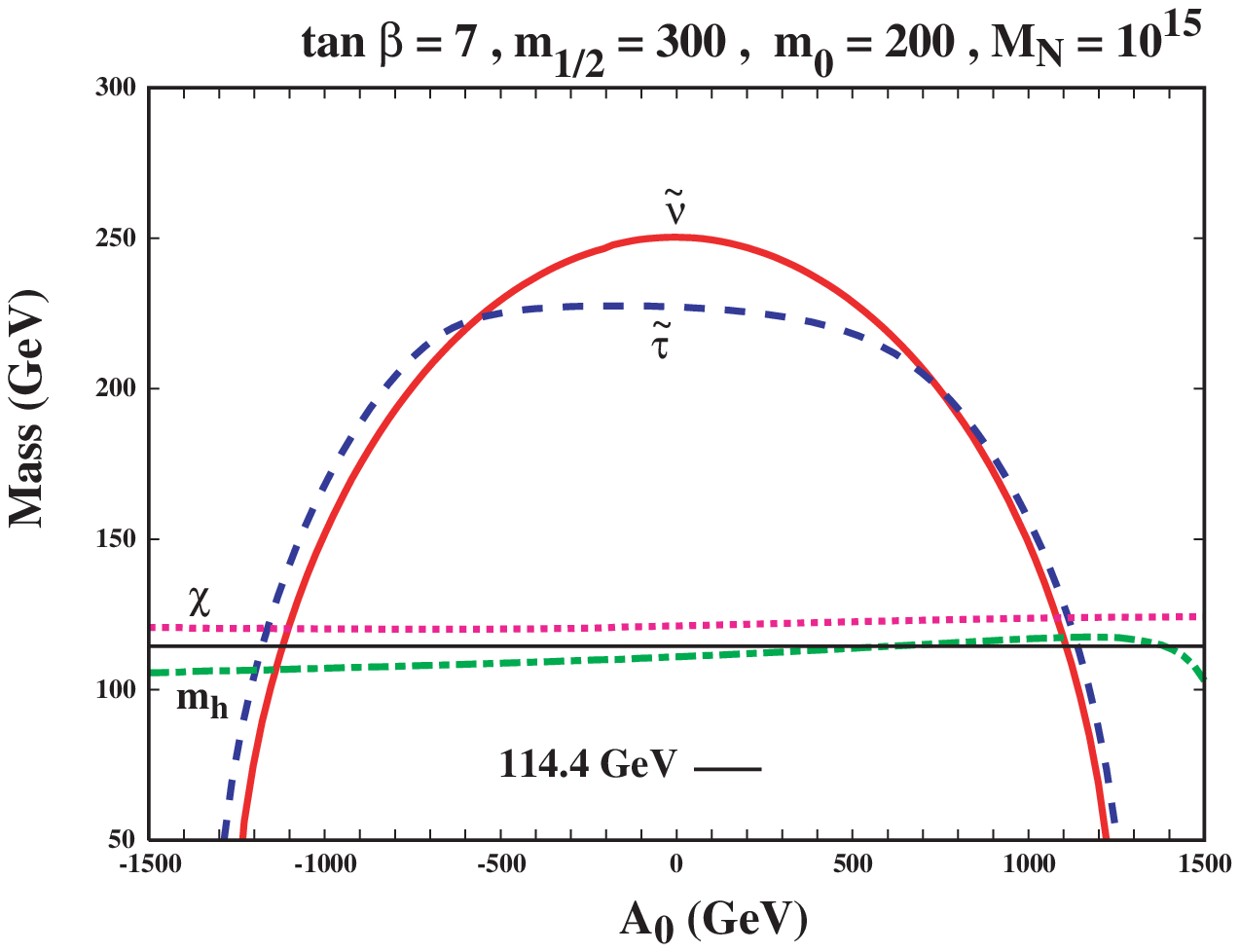}
             }
             }
\caption{\it The mass spectra as a function of $A_0$. 
The CMSSM with $m_0=200$ GeV, $m_{1/2}=300$ GeV, $\tan \beta=7$,  $\mu>0$ is shown in Fig.\ref{am1}. The same mass spectrum is shown in 
Fig.\ref{am11}  for the $\nu$CMSSM with $M_N=10^{15}$ GeV, $m_{\nu}=0.05$ eV. }
     \label{amplot} 
     \end{figure}

We show selected mass spectra for both the CMSSM and  $\nu$CMSSM as a function of $A_0$
with $\tan \beta=7,m_{1/2}=300$ GeV, and $m_0=200$ GeV.
We use a top mass $m_t=172.6$ GeV and a bottom $m_b(m_b)^{MS}=4.25$ GeV in our analysis \cite{cdft}.
In the CMSSM, shown in Fig. \ref{am1}, we see first that the sneutrino mass (heaviest of the masses
shown) is relatively independent of $A_0$ as is the lightest neutralino mass at roughly $m_\chi = 120$ GeV. The lightest stau eigenstate is primarily right-handed and tends to lower values for large $|A_0|$ which then provides a large off-diagonal term in the stau mass matrix. For this choice of CMSSM parameters, the Higgs mass is below the LEP lower bound at $A_0 = 0$ and exceeds the LEP bound for $A_0 \gtrsim 550$ GeV. For reference the LEP bound of 114.4 GeV is
shown by the thin black line. FeynHiggs \cite{fh} was used to calculate $m_h$.

In Fig. \ref{am11}, we show the same set of masses in the $\nu$CMSSM
 with $M_N=10^{15}$ GeV and $m_{\nu}=0.05$ eV.  For these parameters, the neutrino Yukawa coupling is $y_N(Q_{GUT})\sim 2.5$ and $y_N(M_N)\sim 1.8$ when $A_0=1100$ GeV. 
 At $A_0$ close to zero, the sneutrino is heavier than the lighter stau (dominated by its right-handed component) as in the CMSSM. As $|A_0|$ gets large, the composition of the lighter stau
 eigenstate becomes dominated by the left-handed component and it along with the sneutrino
 begin to decrease in mass. When $A_0$ is moderately large we see also that the 
 sneutrino mass falls below the stau mass. 
 This behavior is expected
because $2 y_N^2 X_N$, the featured term in the beta function of
$m_{L_3}^2$ shown in Eq. \ref{rge1}, becomes
large for a large $|A_0|$. 
 The similar behaviors of the sneutrino and stau masses in Fig. \ref{am11} is expected because of their common RGE evolution (when the lighter stau is dominated by  $\tilde {\tau_L}$), with their mass difference coming from the left-right mixings in the stau mass matrix and the D-term contributions due to the sign difference in the SU(2) generator. Once again, the neutralino mass is not very sensitive to $A_0$, and a mass degeneracy between the neutralino and sneutrino masses occurs when $A_0$ is of order the TeV scale and for which the bound $m_h>114.4$ GeV can be satisfied. Therefore, we expect that sneutrino NLSP coannihilation with a neutralino LSP can in principle be realized in such parameter regions.

\begin{figure}[]
\begin{center}    
\epsfxsize = 0.6\textwidth
\epsffile{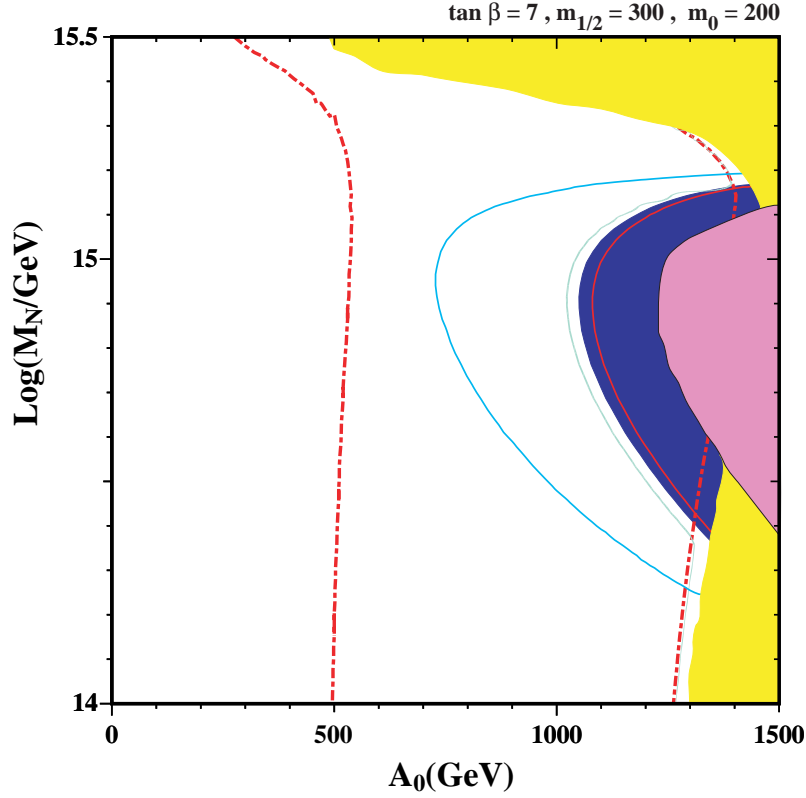}
\end{center}        
\caption{
\label{a0vsmn} {\it The
$A_0-M_N$ plane showing the mass ratios for $m_{\tilde{\nu}}/m_{\tilde{\tau}}=1$ (light blue), $m_{\tilde{\tau}}/m_{\chi}=1$ (red). These ratios decrease towards larger $A_0$. The region with 
 $m_{\tilde{\nu}}/m_{\chi} \le 1$ is shaded dark blue. The thin strip shaded turquoise corresponds
 to the sneutrino coannihilation region where the relic density agrees with the WMAP determination.
The region to the left of the red dot-dashed line (at $A_0 \simeq$ 500 GeV) is excluded by the Higgs mass constraint $m_h>114.4$ GeV.  
There is also a thin region excluded by the Higgs mass to the right of the red dot-dashed curve at 
large $A_0$. 
The remaining shaded regions are excluded because either the stop is lighter than the neutralino (shaded light yellow) or the sneutrino is tachyonic (shaded pink). The parameters are set as $m_{\nu}=0.05$ eV, $m_0=200$ GeV, $m_{1/2}=300$ GeV, $\tan \beta=7$, $\mu>0$.}}
\end{figure}

To gain further insight into the region of parameter space for which sneutrino coannihilation plays
an important role in determining the neutralino relic density, we examine an example of 
a $(A_0,M_N)$ plane. 
Fig. \ref{a0vsmn} is one such example chosen again for our reference point with 
$m_{1/2} = 300$ GeV, $m_0 = 200$ GeV, and $\tan \beta = 7$. The neutrino
mass is fixed at $m_\nu = 0.05$ eV.  This figure shows three contours for mass ratios: To the right of the
light blue contour (inside the loop), $m_{\tilde \nu} < m_{\tilde \tau}$.  Inside the dark blue shaded region, the sneutrino becomes the LSP, and to the right of the red contour (inside the sneutrino LSP region) the stau also becomes lighter than the neutralino. 
We also show the region where the thermal relic abundance  lies in the range
$0.0975<\Omega_{DM} h^2 <0.1223$ corresponding to $2\sigma$ limit determined by WMAP 
 \cite{wmap}, and we can identify the strip lying outside the sneutrino LSP region as the sneutrino coannihilation strip. Also shown are two contours (red dot-dashed) where $m_h = 114.4$ GeV.
 In the region between these curves, the LEP limit is satisfied. 
In the lower right corner (large $A_0$ and low $M_N$) and upper portions of the figure, the stop becomes lighter than the neutralino ( in the light yellow shaded region) and we see a faint stop coannihilation strip to the left of 
this region. Deep inside this region the stop becomes tachyonic.  At the far right,
the sneutrino becomes tachyonic. This region is shaded pink. 
For values of  $M_N$ smaller than that shown, the low-energy mass spectra come to 
resemble those for the CMSSM because the neutrino Yukawa coupling becomes small.

\begin{figure}[]
\begin{center}    
\epsfxsize = 0.48\textwidth
\epsffile{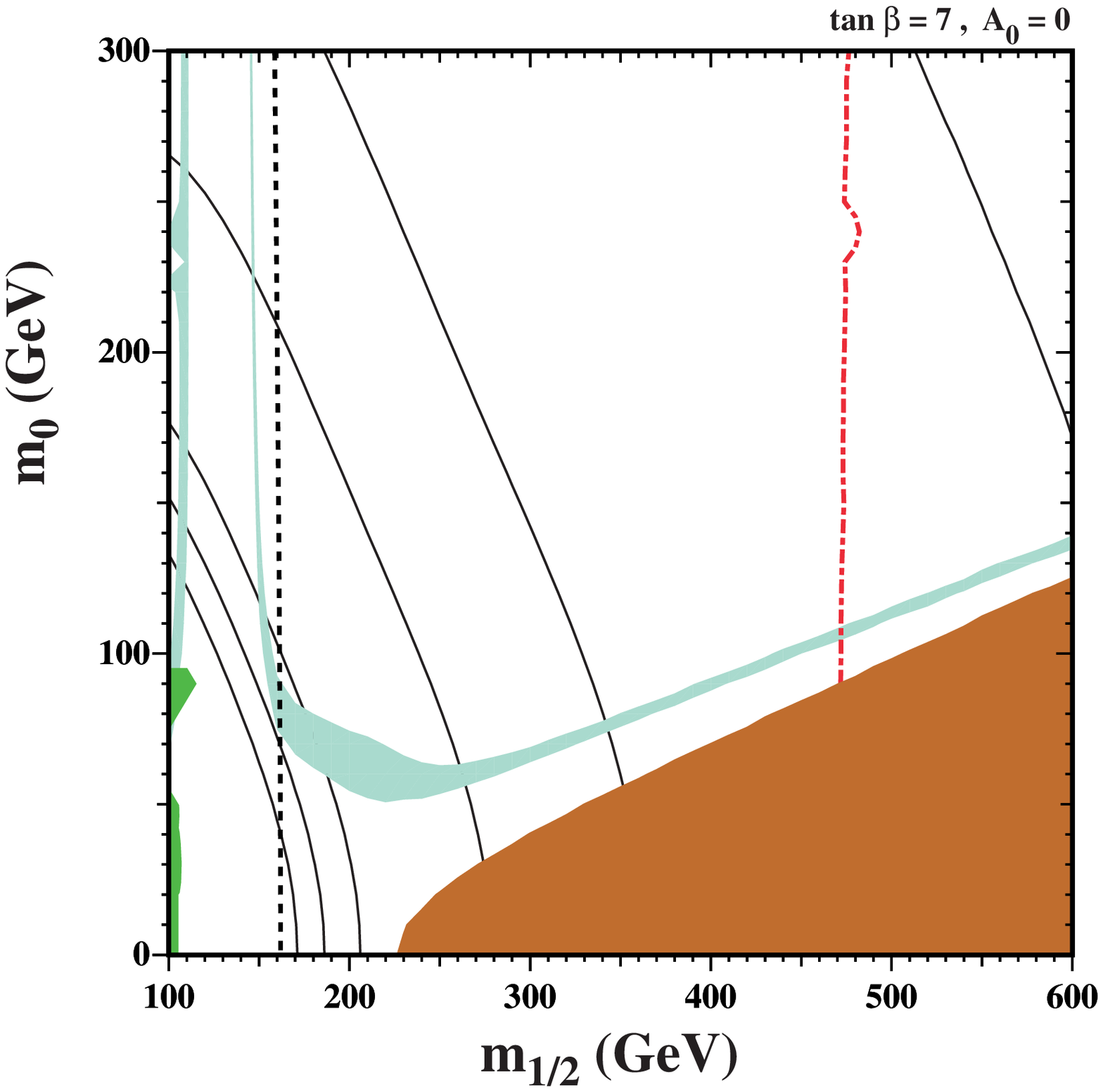}
\epsfxsize = 0.48\textwidth
\epsffile{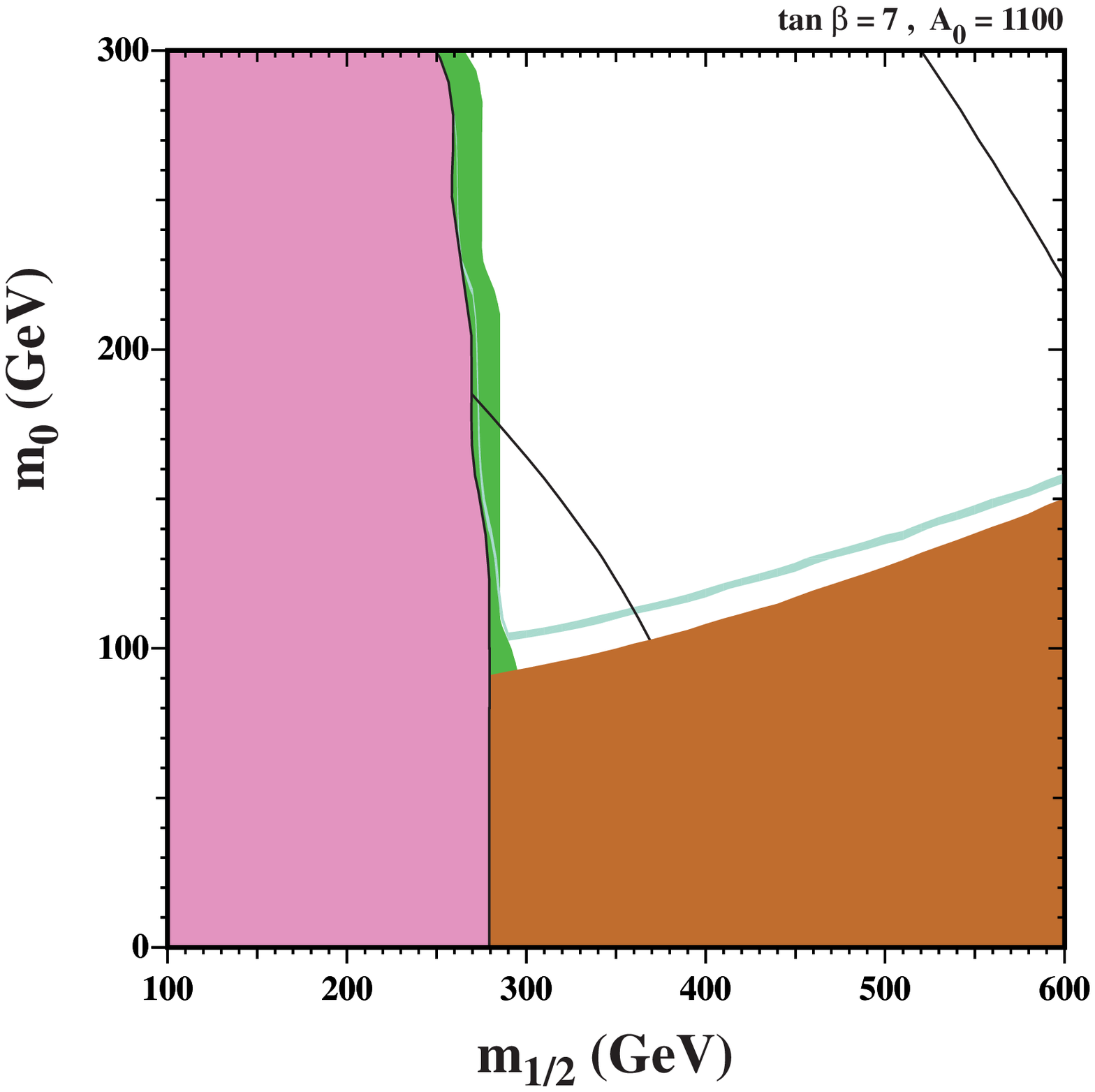}
\end{center}        
\caption{\label{m0vsm12} {\it
The $m_{1/2}-m_0$ plane in the CMSSM with
$\tan \beta=7$ and $A_0=1100$ GeV. The region with the relic density in agreement with WMAP data is 
shown by the turquoise strip corresponding to stau coannihilations. A description of the curves and shaded regions is given in the text.}}
\end{figure}

\begin{figure}[]
\begin{center}    
\epsfxsize = 0.6\textwidth
\epsffile{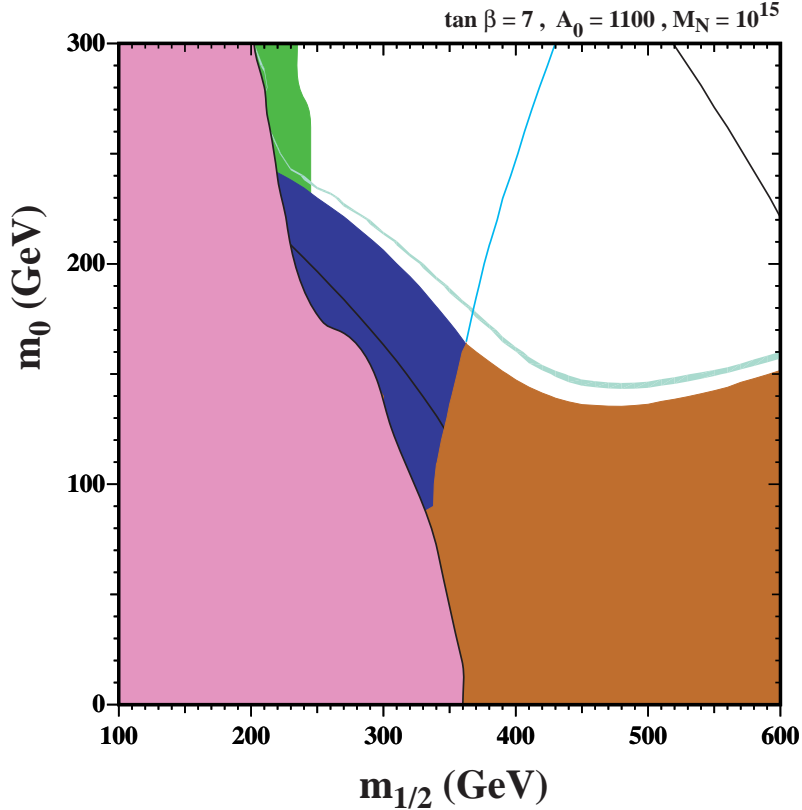}
\end{center}        
\caption{\label{m0vsm12b} {\it The $m_{1/2}-m_0$ plane in the $\nu$CMSSM with
same CMSSM parameter values used in Fig. \ref{m0vsm12} and in addition: $M_N=10^{15}$ GeV, $m_{\nu}=0.05$ eV. The sneutrino is the lightest standard model superparticle in the dark
blue shaded region and at slight higher values of $m_0$, we find the sneutrino co-annihilation region.}
}
\end{figure}

The final set of planes we wish to consider are the $(m_{1/2},m_0)$ planes.
In Figs. \ref{m0vsm12} and \ref{m0vsm12b}, we show examples in both the
CMSSM and  $\nu$CMSSM respectively. In each case, we have fixed $\tan \beta = 7$.  
The case for the CMSSM  is well studied \cite{eoss,stop} and is shown in Fig. \ref{m0vsm12} 
for $A_0 = 0$ and $A_0 = 1100$ GeV. The dark red shaded region at low $m_0$ is also excluded
as there the LSP is the charged stau.  We also see the characteristic stau coannihilation region
which tracks the boundary where $m_\chi = m_{\tilde \tau}$. Here, the relic density agrees with the 
WMAP determination.  In the left panel, we show the chargino mass contour at 104 GeV (black dashed) \cite{lepb} and the Higgs mass contour at 114.4 GeV (red dot-dashed).  Regions to the right of these lines have larger masses. As one can see,
when $A_0 = 0$, the Higgs mass constraint is quite strong and excludes gaugino masses 
$m_{1/2} < 480$ GeV. While the Higgs mass constraint is relaxed at large $A_0$ (as discussed earlier),
for small values of $m_{1/2}$, one of more of the sparticles is tachyonic. This area is shown by the 
pink shaded region in the right panel of Fig. \ref{m0vsm12}.  We also see the good relic density region tend upwards in $m_0$ near the 
tachyonic stop area.  Here, the relic density is controlled by stop coannihilations. This region,
however is excluded by measurements of the $b \to s \gamma$ branching ratio as indicated by the
green shaded strip. 
The area to the right of the green shaded area is consistent at the 2$\sigma$ level with the the constraints coming from the $BR(b \to s \gamma)=(3.55\pm 0.24^{+0.12}_{-0.13})\times 10^{-4}$ \cite{bsg1,bmm} and the current experimental upper bound $BR(B_s \rightarrow \mu^+ \mu^-)<4.7 \times 10^{-8}$ \cite{cdf,bmm}. The LEP constraints on the chargino and Higgs masses only exclude portions of the tachyonic area when $A_0 = 1100$ GeV and are not shown here.
Finally, we have also plotted contours of the anomalous magnetic moment of the muon, $a_{\mu}=(g-2)_{\mu}/2$ which correspond to a
deviation of $a_{\mu}$ from the standard model prediction $(30.2\pm 8.8)\times 
10^{-10}$ \cite{g2}. Shown by the thin black curves are the 1, 2, and 3 $\sigma$ upper and lower bounds.  In the right panel only the 2 and 3 $\sigma$ lower bounds are visible. That is, within the inner curve, $a_\mu$ is within 2$\sigma$ and within the outer curve, it is within $3\sigma$.

When the right handed neutrino is included in our analysis, the ($m_{1/2},m_0$) plane
shows new features in the $\nu$CMSSM as seen in Fig. \ref{m0vsm12b}.
We use the same parameter set and in addition fix $M_N = 10^{15}$ GeV and $m_\nu = 0.05$ eV.
The shadings here are identical to those shown for the CMSSM with the exception of a new
region shown shaded dark blue corresponding to the area where $m_{\tilde \nu} < m_\chi$.
The thin blue line separates regions where the ratio $m_{\tilde \nu}/m_{\tilde \tau}$ is $<1$ (to the left)
and $>1$ (to the right).
Small $m_{1/2}$ is still excluded as sparticles there become tachyonic. 
We see that the stau coannihilation strip is perturbed upwards in $m_0$ where the relic density
becomes controlled by sneutrino coannihilations rather than stau coannihilations. At still smaller
$m_{1/2}$ the relic density strip becomes vertical when stop coannihilation is dominant though
this region is excluded by $b \to s \gamma$. 
Our preferred sneutrino NLSP region is in agreement with the deviation in $a_\mu$ at about the 
2.5$\sigma$ level. Improvements in the accuracy 
of the theoretical predictions as well as the experimental data such as the clarifications of the discrepancies between $e^+ e^-$ and $\tau$ based data would be useful to justify/falsify some of these parameter regions.

\section{Discussion}

In addition to the change of the positions and shapes of the acceptable parameter 
regions in the $\nu$CMSSM compared with those in the CMSSM, we see the stau 
coannihilation region is extended to the sneutrino coannihilation region when the 
sneutrino becomes lighter than the stau. Such regions do not occur in CMSSM 
because the left-handed sneutrino cannot be the NLSP with its mass close to the 
neutralino LSP. One may wonder if the stau coannihilation still plays an important 
role in the sneutrino coannihilation region when the mass of the stau is still 
close to the sneutrino mass $m_{\tilde{\tau}} \sim m_{\tilde{\nu}}> m_{\chi}$. In 
fact, the sneutrino coannihilations, rather than the stau coannihilations, turn
out to be very robust in reducing the LSP neutralino abundance. For instance, let 
us  pick the point $m_0=180$ GeV,$m_{1/2}=365$ GeV in Fig. \ref{m0vsm12b} 
where 
the LSP neutralino abundance is $\Omega_{\chi}h^2=0.107$ with $m_{\tilde{\tau}}=166$ GeV, $m_{\tilde{\nu}}=165$ GeV and $m_{\chi}=152$ GeV. If we artificially shut off sneutrino coannihilation processes in our numerical code,  $\Omega_{\chi} h^2$ increases to $0.350$ while the shutting-off of the stau coannihilation only changes $\Omega_{\chi} h^2$ to $0.126$. 

In addition to the sneutrino coannihilation region with the neutralino LSP we have been discussing, we find a wider parameter region with $\Omega h^2 < 0.104$ at smaller $m_0,m_{1/2}$ where the sneutrino is lighter than the neutralino.  This is the dark blue shaded region in Fig. \ref{m0vsm12b}. A small relic abundance of sneutrinos is expected because sneutrinos can annihilate efficiently through s-wave channels. In this region, however, the sneutrino may be the LSP if there is another source for dark matter \cite{directleft}. On the other hand, the sneutrino may be the NLSP if we assume that the gravitino is the LSP.  In this case, the sneutrino will decay into
gravitinos and the gravitino dark matter abundance $\Omega_{\tilde{G}}h^2$ now becomes   
\ba
\Omega_{\tilde{G}}h^2=\frac{m_{\tilde{G}}}{m_{\tilde{\nu}}}\Omega_{\tilde{\nu}}h^2+\Omega^T_{\tilde{G}}h^2
\ea
In addition to the first term representing the contribution from the decay of the NLSP sneutrino, there is a thermal gravitino relic abundance contribution $\Omega^T_{\tilde{G}}h^2$ originating from the thermal production during the reheating era after inflation which is heavily model dependent\footnote{We refer the reader to the previous literature for the cosmological constraints coming from Big Bang Nucleosynthesis and the erasure of the small scale structure at the gravitino decoupling era as those constraints are analogously applicable to the $\nu$CMSSM \cite{covi2,john2,matsumoto}.}.  

Similar regions with a sneutrino NLSP and gravitino LSP can be found in the NUHM \cite{john2}.
However, there is an important difference between those regions and the ones
we are discussing in the $\nu$CMSSM.  In the NUHM, the sneutrino can become the lightest
standard model superpartner when $|\mu|$ is large.  As discussed earlier, this occurs when
$S$ is large and negative. This often results in soft Higgs masses $m_{{H_u},{H_d}}^2 < 0$
at the GUT scale. This may lead to problems of vacuum stability at low energies \cite{gut}. 
This is never a problem in the $\nu$CMSSM.

Finally, let us briefly comment on the prospects for the collider signals for the sneutrino NLSP scenarios. The sneutrino can be directly produced, for instance, via the off-shell gauge bosons $q\bar{q} \rightarrow Z^0 \rightarrow \tilde{\nu} \tilde{\nu}^*, q\bar{q}' \rightarrow W^{\pm} \rightarrow \tilde{\nu} \tilde{l}^*, \tilde{\nu}^* \tilde{l}$ at the LHC or $e^+ e^-\rightarrow Z^0 \rightarrow \tilde{\nu} \tilde{\nu}^*$ at the ILC \cite{ilc}. 
The NLSP sneutrino however decays invisibly into the LSP neutralino $\chi$ via $\tilde{\nu} 
\rightarrow \nu \chi$ with the decay rate of 
\ba
\Gamma_{\tilde{\nu} \rightarrow \chi \nu }
=
\frac{g^2}{16\pi}|X^{\nu}_{\chi}|^2
m_{\tilde{\nu}}\left(  1-\frac{m^2_{\chi}}{m_{\tilde{\nu}}}     \right)^2 
\ea
where $X^{\nu}_{\chi}=(N_{11}\tan \theta_W-N_{12})/\sqrt2$ ($N$ is the neutralino 
mixing matrix in the $(\tilde{B},\tilde{W^0},\tilde{H_d^0},\tilde{H_u^0})$ basis). 
In addition to this invisible two-body decay of $\tilde{\nu}$, there are also 4-body final 
state decay channels such as $\tilde{\nu} \rightarrow 
l^+ \chi \bar{f} f'$ ($\bar{f} f'=\bar{\nu} l^-, \bar{u}d$). The production of a sneutrino
via a virtual $W$ at the LHC is followed by 
$\tilde{l} \rightarrow l \chi$, so that an isolated hard charged lepton with missing energy can result even though the large background from the direct $W$ decay is problematic to distinguish this signal \cite{bae3}. A potentially more promising process can be the search of sneutrinos from two-body leptonic decays of the pair-produced charginos in $e^+ e^-$ collisions at the ILC, which could possibly give a characteristic decay lepton energy spectra with the controllable backgrounds \cite{acce,kal}.

For the gravitino LSP, on the other hand, the NLSP sneutrino decays into the LSP gravitino $\tilde{G}$ with a decay rate
\ba
\Gamma_{\tilde{\nu} \rightarrow  \tilde{G} \nu}
=
\frac{1}{48\pi}\frac{m^5_{\tilde{\nu}}}{M_{pl}^2 m^2_{\tilde{G}}}
\left( 1-\frac{m^2 _{\tilde{G}}}{m^2_{\tilde{\nu}}}  \right)^4
\ea
In these gravitino LSP scenarios, the sneutrino NLSP typically would not decay within the 
detector for the parameter range of interest. Collider signatures are imprinted in 
the decays from the heavier states into the sneutrino, and the decays of the heavier 
neutralinos and charginos into the sneutrino have been studied \cite{john2,gje,but,laura,kra}. 
The detailed 
study of the collider signatures for the sneutrino NLSP in the $\nu$CMSSM however depend on 
the neutrino mass structure which we did not explore in this letter. It would be of great interest to see how such accelerator signals can be related to the GUT scale seesaw mechanism and is left for future work.

We have shown that the GUT scale seesaw mechanism can play an important role in low-energy phenomenology, and in particular the properties of dark matter. Even though a 
right-handed neutrino is integrated out at its heavy mass scale close to the GUT scale, it can significantly affect the RGE evolution of the mass spectra due to a large neutrino Yukawa coupling. We examined the  `sneutrino coannihilation region'  in which we obtain the correct relic density 
of a neutralino LSP with a left-handed sneutrino NLSP.

\subsection*{Acknowledgments}    
We thank G. Kane, A. Pierce and J. Shao for useful discussions.
The work of K.K. was supported in part by Michigan Center for Theoretical Physics . The work of K.K. and K.A.O was supported in part by DOE grant DE-FG02-94ER-40823 and the William I. Fine Theoretical Physics Institute.


\end{document}